\documentclass[%
 reprint,
 amsmath,amssymb,
 aps,
]{revtex4-1}

\usepackage{graphicx}
\usepackage{dcolumn}
\usepackage{bm}

\newcommand{\aap}{A.\&Ap.}

\newcommand{\mnras}{MNRAS}

\newcommand{\jcap}{JCAP}

\newcommand{\apss}{Ap\&SS}

 \newcommand{\mbf}[1]{\mbox{\boldmath ${#1}$}}

\usepackage{color}

\newcommand{\revi}[1]{#1}

\begin{document}

\preprint{APS/123-QED}

\title{
\revi{Differentiating Warm Dark Matter Models through 21cm Line Intensity Mapping: A Convolutional Neural Network Approach}}

\author{Koya Murakami$^1$}
\email{koya.murakami9627@gmail.com}
\author{Kenji Kadota$^{2,3}$}
\email{kadota@ucas.ac.cn}
\author{Atsushi J. Nishizawa$^{4,5,6}$}
\email{atsushi.nishizawa@iar.nagoya-u.ac.jp}
\author{Kentaro Nagamine$^{7,8,9,10}$}
\email{kn@astro-osaka.jp}
\author{Ikkoh Shimizu$^{11}$}
\email{ishimizu@sg-u.ac.jp}

\affiliation{
$^1$Department of Physics, Nagoya University, Furocho, Chikusa, Nagoya, 464-8602, Japan\\
$^2$School of Fundamental Physics and Mathematical Sciences, Hangzhou Institute for Advanced Study, University of Chinese Academy of Sciences (HIAS-UCAS), Hangzhou 310024,
China\\
$^3$International Centre for Theoretical Physics Asia-Pacific (ICTP-AP), Beijing/Hangzhou,
China\\
$^4$DX Center, Gifu Shotoku Gakuen University, 1-1 Takakuwanishi, Yanaizucho, Gifu, 501-6194, Japan \\
$^5$Institute for Advanced Research, Nagoya University, Furocho, Chikusa, Nagoya, Aichi, 464-8602, Japan\\
$^6$Kobayashi Maskawa Institute, Nagoya University, Furocho, Chikusa, Nagoya, Aichi, 464-8602, Japan\\
$^7$Theoretical Astrophysics, Department of Earth and Space Science, Graduate School of Science, Osaka University, 1-1 Machikaneyama, Toyonaka, Osaka 560-0043, Japan\\
$^8$Theoretical Joint Research, Forefront Research Center, Graduate School of Science, Osaka University, 1-1 Machikaneyama, Toyonaka, Osaka 560-0043, Japan\\
$^9$Kavli IPMU (WPI), The University of Tokyo, 5-1-5 Kashiwanoha, Kashiwa, Chiba, 277-8583, Japan\\
$^{10}$Department of Physics \& Astronomy, University of Nevada, Las Vegas, 4505 S. Maryland Pkwy, Las Vegas, Nevada 89154-4002, USA\\
$^{11}$Shikoku Gakuin University, 3-2-1 Bunkyocho, Zentsuji, Kagawa, 765-8505, Japan
}



\date{\today}

\begin{abstract}
We apply the convolutional neural networks (CNNs) to the mock 21cm maps from the post-reionization epoch to show that the $\Lambda$ cold dark matter and warm dark matter (WDM) model can be distinguished for WDM particle masses $m_{FD}<3$\,keV, under the assumption of thermal production of WDM following the Fermi-Dirac (FD) distribution. 
We demonstrate that the CNN is a potent tool in distinguishing the dark matter masses, highlighting its sensitivity to the subtle differences in the 21cm maps produced by varying dark matter masses.
Furthermore, we extend our analysis to encompass different WDM production mechanisms, recognizing that the dark matter production mechanism in the early Universe is among the sources of the most significant uncertainty for the dark matter model building.

In this work, given the mass of the dark matter, we discuss the feasibility of discriminating four different WDM models: Fermi-Dirac (FD) distribution model, neutrino minimal Standard Model ($\nu$MSM), Dodelson-Widrow (DW), and Shi-Fuller (SF) model. For instance, when the WDM mass is 2\,keV, we show that one can differentiate between CDM, FD, $\nu$MSM, and DW models while discerning between the DW and SF models turns out to be challenging. Our results reinforce the viability of the CNN as a robust analysis for 21cm maps and shed light on its potential to unravel the features associated with different dark matter production mechanisms.

\end{abstract}

\maketitle


\section{\label{sec:introduction}introduction}
%

The $\Lambda$ cold dark matter ($\Lambda$CDM) model has garnered widespread acceptance within the cosmological community due to its robust capability to account for a multitude of observational phenomena. Central to this model is the presumption of cold dark matter (CDM), posited to consist of nonrelativistic particles. 
Despite its remarkable success and the extensive insights it provides into the structure and dynamics of the universe, the $\Lambda$CDM model falls short of fully elucidating the intricate nature and characteristics of dark matter. Key aspects such as its production mechanisms and the evolution of small-scale structures remain among the most profound mysteries in cosmology, and the possibility of a dark matter model that goes beyond the simple cold dark matter paradigm presents an intriguing avenue for exploration.

The warm dark matter (WDM) model stands as a prominent alternative candidate among various dark matter models. We explore the thermally produced WDM, distinguished by a phase space distribution governed by Fermi-Dirac statistics (referred to as ``FD" in subsequent discussions). At the time of production, WDM particles are relativistic, and their free-streaming suppresses the growth of matter fluctuation in the universe on small scales. The extent of this free-streaming, and consequently the suppression effect, is characterized by the WDM mass. Consequently, by examining its impact on the formation and evolution of cosmic structures, we can investigate the mass range of WDM.

The cosmological observation gives some constraints or their forecasts, with the Lyman-$\alpha$ forest as a prime example. 
It traces the matter distribution of the universe along the line of sight, delivering constraints on the particle mass of the FD warm dark matter model. 
Specifically, the analysis of the power spectrum of Lyman-$\alpha$ forest sets a lower limit on the FD particle mass to be of the order of $\mathcal{O}(1)$ keV \citep{Viel2013, Garzilli2019, Garzilli2019a, 2023PhRvD.108b3502V}.  

The 21 cm signals, emitted from the neutral hydrogen (HI), offer another avenue for probing the distribution of matter in the universe. 
These signals are the focus of ongoing and planned observations by several major radio astronomy projects, including the
Murchison Wide-field Array (MWA) \citep{MWA}, Canadian Hydrogen Intensity Mapping Experiment (CHIME) \citep{CHIME}, Hydrogen Intensity and Real-time Analysis eXperiment (HIRAX) \citep{HIRAX}, and Square Kilometer Array (SKA) \citep{SKA}. Ref.~\citep{2015JCAP...07..047C} utilizes the power spectrum of 21 cm signals, specifically anticipating the SKA data, to forecast constraints on the WDM particle mass. 
These efforts highlight the critical role of radio observations in advancing our understanding of dark matter and the larger cosmic structure.

In the context of analyzing large-scale cosmic structures, we leverage image-based analysis by utilizing image-based analysis with convolutional neural networks (CNNs). CNNs adeptly extract information on matter distribution directly from images, offering the potential for more stringent constraints on the WDM mass than traditional methods such as the power spectrum. Ref.~\cite {2024MNRAS.527..739R} demonstrates that  CNNs can yield more accurate predictions of WDM mass by analyzing dark matter maps generated from N-body simulations. Similarly, our prior work \citep{ourwork} also shows that the CNN analysis is superior to power spectrum analysis in extracting detailed information from maps of 21 cm signals derived from cosmological hydrodynamic simulations.

Firstly, this work considers the binary classification of CDM and FD models with different masses. We generate images of the 21 cm signals from hydrodynamic simulations across a range of redshifts $z=[3,4,5,6]$ during the post-reionization era.  We then proceed to compare the effectiveness of CNNs against traditional power spectrum methods in performing these 
classifications.
In addition, we explore implementing a CNN architecture designed to simultaneously incorporate data from multiple redshifts, aiming to assess whether this multiredshift approach yields superior performance compared to analyses restricted to a single redshift. 
Additionally, we consider the potential impact of observational noise, as encountered in SKA-Low surveys, and present the outcomes of binary classifications under both single and joint redshift analysis. 

As a further application of our analysis, we study if future 21cm observations can distinguish different dark matter production mechanisms.
Here, we consider the scenarios where the dark matter mass is predetermined, possibly from collider and direct/indirect dark matter search experiments, and apply our CNN analysis to the mock 21cm map data for the specified WDM mass with different WDM production scenarios.

For instance, accelerator experiments generate dark matter through particle collisions, and the relevant dark matter abundance is independent of the total matter abundance in the Universe $\Omega_m$. These experiments can elucidate dark matter properties, including its mass and interaction cross-sections, by analyzing aspects of dark matter kinematics, such as transverse missing energy.
Consequently, it is plausible that such alternative dark matter experiments could significantly narrow down the constraints on dark matter mass, offering a complementary perspective to the cosmological surveys. 
This potential for cross-disciplinary synergy would motivate our approach, aiming to highlight the connection between our cosmological studies and other dark matter search experiments.

We study, for illustration, four different WDM production mechanisms.
Besides FD, which possesses thermal phase space distributions, we study other commonly discussed nonthermal production mechanisms $\nu$MSM (neutrino minimal standard model), DW (Dodelson-Widrow), and SF (Shi-Fuller).
In DW mechanism \citep{Dodelson1994}, the sterile neutrino dark matter is produced through the oscillations with the active neutrinos in the cosmic plasma. The oscillations convert a fraction of an active neutrino into a long-lived sterile neutrino, acting as dark matter. 
In SF mechanism \citep{Shi1999}, sterile neutrino dark matter production is significantly amplified compared to the DW mechanism. The enhancement occurs due to resonant mixing between active and sterile neutrinos in the presence of a strong lepton asymmetry. The SF mechanism addresses some of the observational constraints on the DW mechanism by allowing for a smaller mixing angle due to the resonant enhancement.
The $\nu$MSM \citep{Shaposhnikov2006} extends the Standard Model by including additional sterile neutrinos, which can mix with the active neutrinos, similar to the DW mechanism. However, $\nu$MSM distinguishes itself by evading some of the parameter constraints faced by the DW mechanism. The lightest of these sterile neutrinos acts as the WDM candidate. The model is appealing as it not only provides a viable WDM candidate but also addresses other cosmological and particle physics phenomena, such as the baryon asymmetry of the universe and the masses of active neutrinos. 

While some WDM production mechanisms, such as those produced thermally, can lead to the Fermi-Dirac distributions, others do not possess such simple thermal phase distributions. We, however, point out that the main characteristics that the cosmological observables are sensitive to are the free streaming scale and the relic abundance, and there are degeneracies among those different production mechanisms. One common practice in the literature is the mapping between 
the mass for the Dodelson Wildrow scenario and that for the thermally produced warm dark matter, by demanding their free streaming scale and abundance match, which can allow a reinterpretation of mass constraints from large-scale structure data across various WDM models \citep{1996ApJ...458....1C}.
Similarly, we can rescale the masses among the different aforementioned production scenarios, and we, for concreteness, adopt the following mapping \citep{de2012cosmological, Destri2013,deVega:2011gg}, 
 \begin{align}
     m_{\rm DW} &\sim 2.85\ \mathrm{keV} \left(\frac{m_{\rm FD}}{\mathrm{keV}} \right)^{4/3} , \label{eq:FD-DW} \\ 
     m_{\rm SF} &\sim 2.55\ m_{\rm FD} , \label{eq:FD-SF} \\
     m_{\nu \mathrm{MSM}} &\sim 1.9\ m_{\rm FD} ,
     \label{eq:FD-nuMSM}
 \end{align}
 where $m_{X}$ represents the dark matter mass of the model $X$.
 Even though those nonthermal production mechanisms do not have the Fermi-Dirac phase space distributions, the Fermi-Dirac phase space distributions with these rescaled masses effectively replicate the identical matter power spectra as those nonthermal distributions with $m_{WD}$ due to the degeneracies in the cosmological observables (notably the matter fluctuation suppression scales and amplitudes). 
The advantage of this degeneracy represented by such a mass rescaling is that we can apply our methodology and techniques to study the Fermi-Dirac distributions to our setup for different production scenarios with the common mass.

In this work, we also apply CNN for the classification between the WDM models for the mock 21 cm map. We discuss different production mechanisms for a predetermined mass and demonstrate the potential of CNN to constrain the dark matter model through large-scale structure analysis.

This paper is organized as follows. In Sec. \ref{sec:wdm_models}, we show the calculation of the power spectrum of the WDM for the FD model. Sec. \ref{sefc:dataset} describes the configuration of our simulation and how to generate images for CNN. In section \ref{sec:method}, we show the architecture of CNN and the neural network for image-based and power spectrum analysis, respectively. Sec. \ref{sec:result} shows the binary classification results between the CDM and FD models and the classification of WDM models. Finally, we summarize our work in section \ref{sec:summary}.

%
\section{warm dark matter models}
\subsection{\label{sec:wdm_models} Fermi-Dirac WDM model}
%

\begin{figure}
    \centering
    \includegraphics[width=\linewidth]{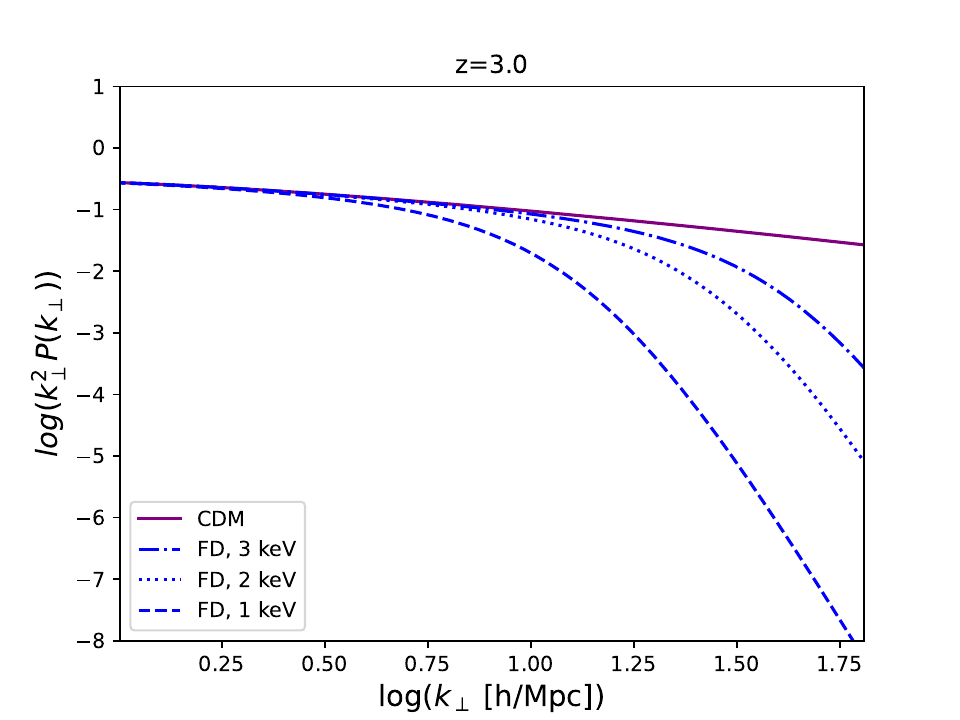}
    \caption{
     The projected 2D power spectra for CDM (solid), 1 (dashed), 2 (dotted), and 3 (dashed-dotted) keV FD WDM at $z=3$. The projection length is $25\ h^{-1}$Mpc.
    }
    \label{fig:pk_wdm_mass}
\end{figure}

First, we consider the fermion WDM for which the phase space distribution follows the Fermi-Dirac distribution, written as
\begin{equation}
\label{eq:psd_wdm}
f(p) = \frac{1}{\exp \left( p / (\alpha T_\gamma) \right) + 1},
\end{equation}
where $p$ is the momentum, $T_\gamma$ is the photon temperature, and $\alpha$ is the parameter representing the ratio of the dark matter temperature to the photon temperature. We refer to this WDM as the FD (Fermi-Dirac).
The dark matter abundance $\Omega_{\rm DM}$, the dark matter particle mass $m_{\rm FD}$, and $\alpha$ are related as
\begin{equation}
\label{eq:relation_wdm_params}
\Omega_{\rm DM} = \frac{\alpha^3}{4/11} \frac{m_{\rm FD}}{94\ \mathrm{eV}}.
\end{equation}
The FD particles are produced as relativistic particles, resulting in suppressed clustering at small scales within their free-streaming scale, depending on their particle mass.

We use \texttt{CLASS}\cite{Lesgourgues2011a} as the Boltzmann solver and compute the matter power spectrum for the CDM model. And then, we define the ratio of the FD and CDM power spectrum as 
\begin{equation}
\label{eq:ratio_wc}
T^2 (k) \equiv \frac{P_{\rm FD} (k)}{P_{\rm CDM}(k)},
\end{equation}
where $P_{\rm FD} (k)$ and $P_{\rm CDM} (k)$ is the matter power spectrum for the FD and CDM model, respectively, and $k$ is the wave number. We can calculate $T^2 (k)$ as \citep{Destri2013}
\begin{equation}
    T^2 (k) = \Phi \left( \frac{k}{k_{1/2}} \right),
\end{equation}
where $k_{1/2}$ is the wave number satisfying $T^2 (k_{1/2}) = 1/2$ and described using the FD particle mass $m_{\rm FD}$ as
\begin{equation}
\label{eq:k_half}
  k_{1/2} = 6.72 \left( \frac{m_{\rm FD}}{\mathrm{keV}} \right)^{1.12} h \mathrm{Mpc}^{-1}.
\end{equation}
$\Phi$ is the analytic formula and is written as 
\begin{equation}
\label{eq:Phi-for-Transfer}
    \Phi (x) = \frac{1}{[1 + (2^{1/b} - 1)x^a]^b},
\end{equation}
where $a$ is $=1.304$, and $b$ is $= 4.478$. Figure~\ref{fig:pk_wdm_mass} shows the dimensionless power spectra on the projected 2D plane for CDM and FD with the masses 1, 2, and 3 keV. The lighter FD WDM has a larger free-streaming scale, and we can see the suppression of the amplitude of the power spectrum at a larger scale for the lighter FD WDM. 

\subsection{\label{sec:other_wdm_models} Other WDM models}

\begin{figure}
    \centering
    \includegraphics[width=\linewidth]{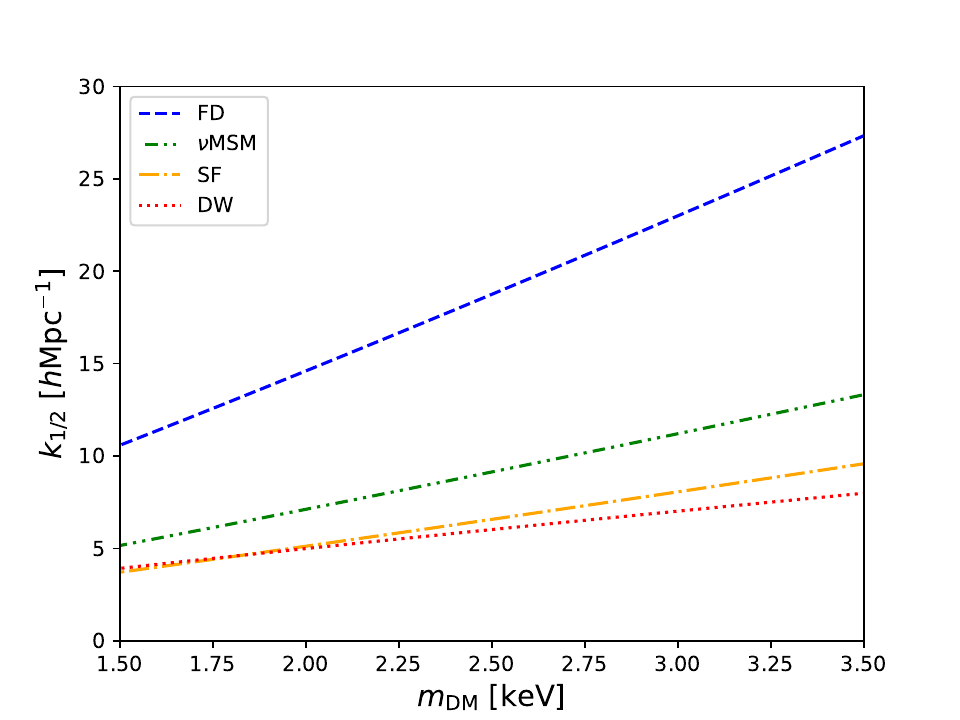}
    \caption{
    The relation between the particle mass of dark matter and $k_{1/2}$. Each line respectively corresponds to the FD (blue, dashed), $\nu$MSM (green, dash-dot-dot), SF (orange, dash-dotted), and DW (red, dotted) model.
    }
    \label{fig:k_half}
\end{figure}

\begin{figure}
    \centering
    \includegraphics[width=\linewidth]{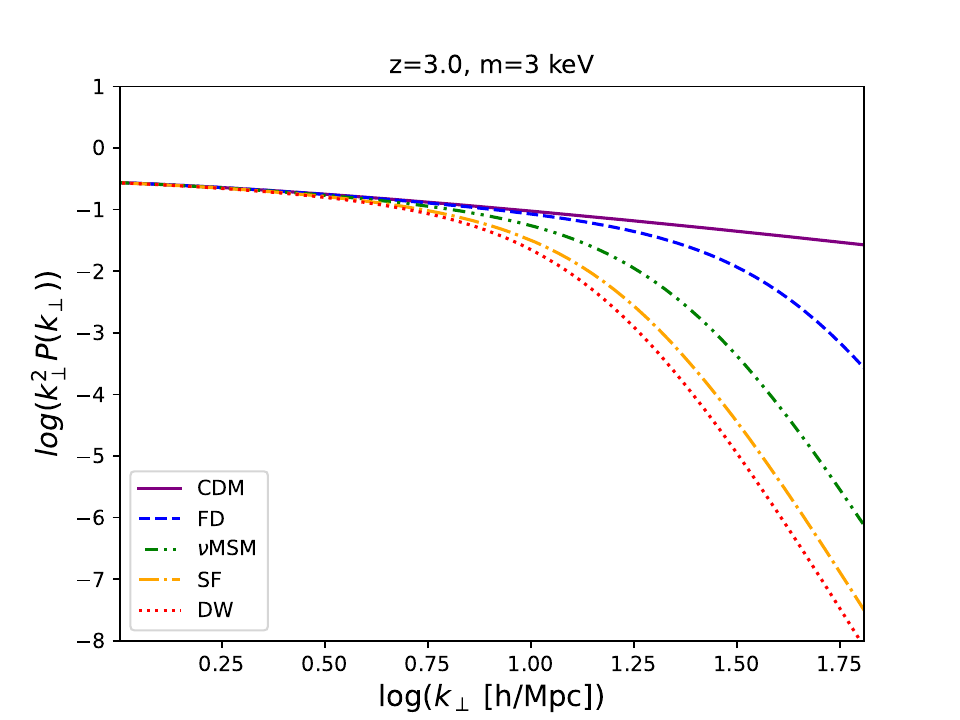}
    \caption{
     The projected 2D linear power spectra for the CDM (purple, solid), FD (blue, dashed), $\nu$MSM (green, dash-dot-dot), SF (orange, dash-dotted), and DW (red, dotted) WDM at $z=3$. Here, we assume the particle mass of WDM is 3 keV. The projection length is $25\ h^{-1}$Mpc.
    }
    \label{fig:pk_wdm_model}
\end{figure}

This work considers the additional three WDM models mentioned in Sec.~\ref{sec:introduction}: $\nu$MSM, SF, and DW. The production mechanisms for these models differ from the thermal production represented by the FD model, and the phase space distributions and, consequently, the free streaming scales differ. We, however, simplify our analysis by calculating the power spectra for these WDM models using FD distributions through the mass mapping given by Eq.~(\ref{eq:FD-DW})$-$(\ref{eq:FD-nuMSM}) in Sec.~\ref{sec:introduction}, which suffices for our purpose of demonstrating the potential power of the 21 cm observations to discriminate among different dark matter production mechanisms.

We analyze the power spectra for a given WDM mass. The dark matter mass can be determined from other complementary experiments, such as a collider experiment, where the properties of dark matter, such as their momentum, are obtained from the kinematics. 
For a given WDM mass, we then reinterpret its mass as the mass of the FD by using Eq.~(\ref{eq:FD-DW})-(\ref{eq:FD-nuMSM}). Finally, we can calculate $k_{1/2}$ and the power spectra for the $\nu$MSM, SF, and DW following the procedure for the FD model by the reinterpreted mass. 
Figure~\ref{fig:pk_wdm_model} and Figure~\ref{fig:k_half} show the projected linear power spectrum at $z=3$ for the WDM particle mass $m_{\rm DM} = 3$ keV and $k_{1/2}$ for each WDM model. As we can see in these figures, $k_{1/2}$ and the power spectra are different between the WDM models even when their particle masses are the same due to the different free-streaming scales.

%
\section{\label{sec:dataset} Data}
%

\begin{figure*}
  \centering
  \includegraphics[keepaspectratio,width=\linewidth]{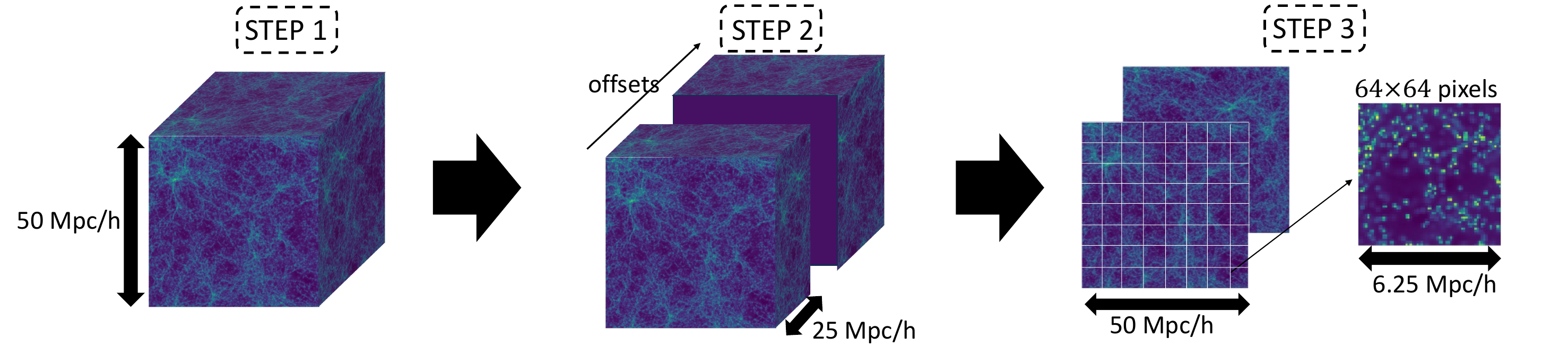}
  \caption{
  The illustration of the procedure of making images from the simulation.
  \revi{STEP1: Define the physical quantities on the regular grid of 512$^3$ using SPH kernel. STEP2: Project them along the LoS over 25 Mpc/$h$ width with the offsets of $\sim 0.1$Mpc$/h$ steps for the data augmentation. STEP3: The image is subdivided into $8\times 8$ images, and the $\delta T_b$ image is transformed with $sinh^{-1}$ function to increase the dynamic range. Please see the text (in Sec. \ref{ssec:images})for more detailed procedures.}
  }
  \label{fig:making_images}
\end{figure*}

\begin{figure*}
  \vspace{-1in}
  \includegraphics[keepaspectratio,width=\linewidth]{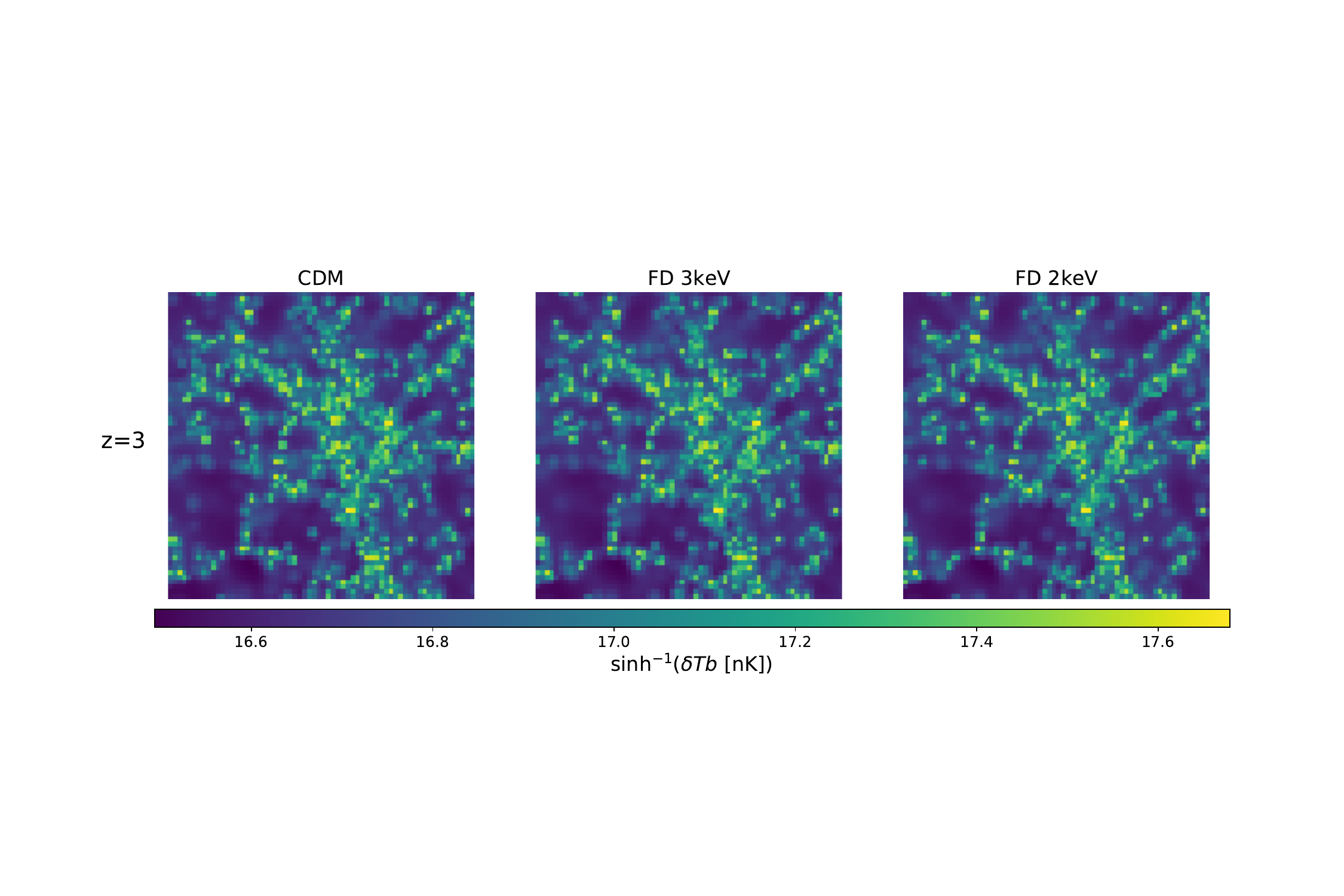} \\ \vspace{-2.4in}
  \includegraphics[keepaspectratio,width=\linewidth]{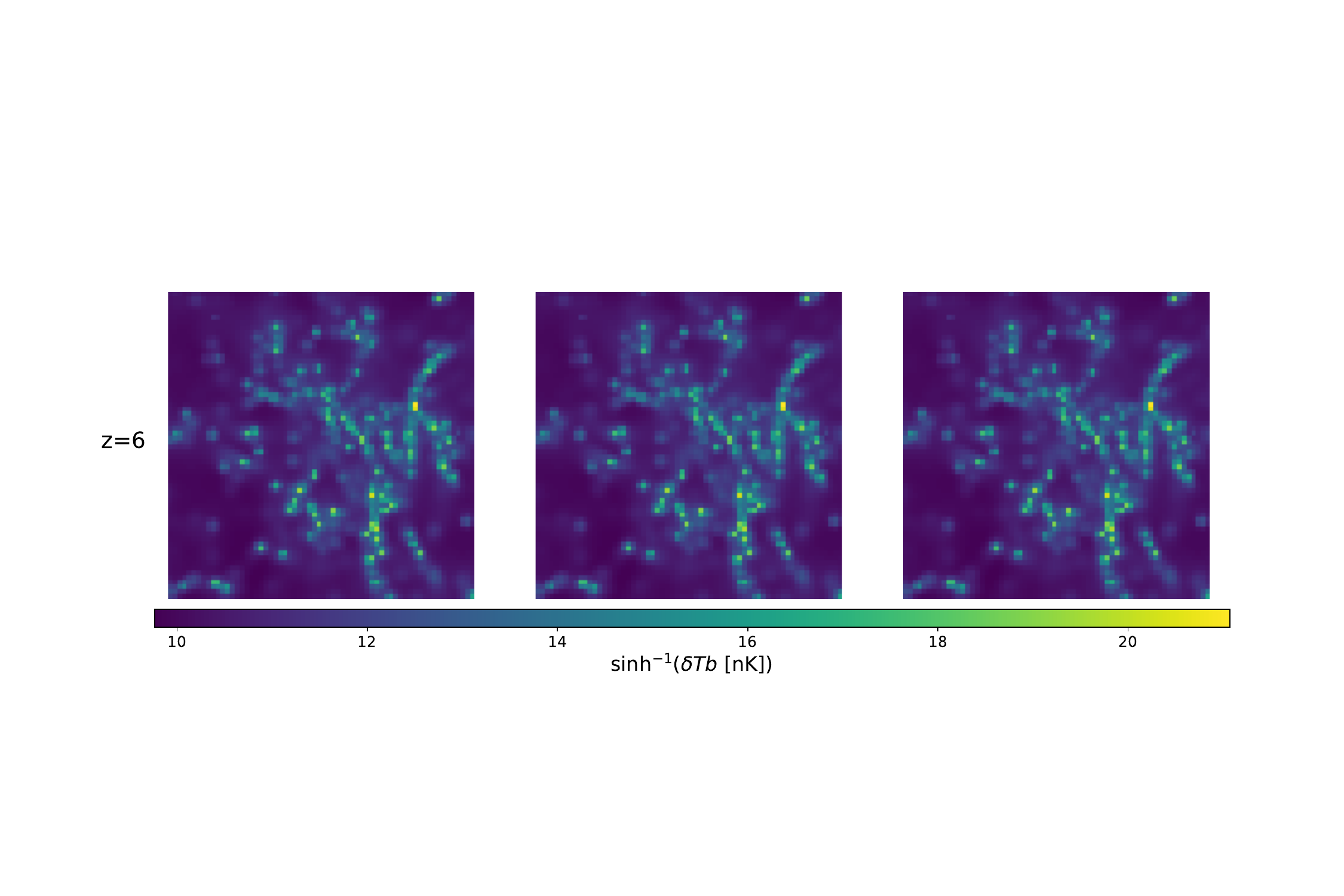}
  \vspace{-1.2in}
  \caption{
  These are examples of the image for CDM (left), 3 keV FD WDM (middle), and 2 keV FD WDM (right) at $z=3$ (upper) and  $z=6$ (lower). Each image includes the $6.25 \times 6.25\ [(h^{-1} \mathrm{Mpc})^2] \times 25\ [h^{-1} \mathrm{Mpc}]$ region of a simulation box and has $64 \times 64$ pixels. These images come from the same region of each simulation box.
  }
  \label{fig:eg_images}
\end{figure*}

\subsection{\label{ssec:simulation}Gadget-Osaka simulation}

In this work, we perform hydrodynamic simulations to generate the mock 21 cm intensity map for the different FD masses and the different WDM models. We assume the cosmological parameter obtained by Planck satellite \cite{Planck2018}; $\Omega_m = 0.3111$, $\Omega_\Lambda = 0.6889$, $\Omega_b = 0.049$, $h = 0.677$, and $\ln 10^{10} A_s = 3.047$. 
First, we consider the standard CDM and FD models with different masses.

The initial power spectrum for the CDM model at $z=99$ is calculated using the \texttt{CLASS}. For FD models, we compute its initial power spectrum following the formulae in Sec.~\ref{sec:wdm_models}. We generate the initial conditions from these input power spectra with \texttt{2LPTic} \cite{Crocce2006}. In making the initial condition, we apply the glass realization to remove the grid pattern in particle distribution and to avoid unrealistic features in the matter distribution \citep{Gotz2002, Gotz2003}.

We use \texttt{GADGET3-Osaka} \citep{10.1093/mnras/stw3061, 10.1093/mnras/stz098} to solve the evolution of matter distribution caused by the gravitational interaction and gas physics. It is the cosmological smoothed hydrodynamic (SPH) code based on \texttt{GADGET3} \cite{Springel2005}. Our simulations have $256^3$ dark matter particles and $256^3$ gas particles in a box whose comoving size is $50\ h^{-1}$Mpc on a side. Therefore, the dark matter particle's mass in our simulation is $5.4 \times 10^8\ M_\odot / h$, where $M_\odot$ is the solar mass. We generate the initial condition at $z=99$ and stop the simulation at $z=3$. This simulation includes the star formation, supernova feedback, UV radiation background, and radiative heating and cooling. The cooling effect is solved by Grackle chemistry and cooling library\cite{grackle}. For the star formation model, we apply the model used in the AGORA project\citep{Kim_2014, Kim_2016}, the supernova feedback model described in \cite{10.1093/mnras/stz098}, and the uniform UV radiation background \cite{Haardt_2012}. We do not consider the self-shielding of HI gas, which is the obstruction of UV radiation by thick HI gas.

Throughout this work, we use the snapshots at $z = [3,4,5,6]$. This redshift range corresponds to the range observed by the SKA-Low survey after the post-reionization epoch.

\subsection{\label{ssec:images} Images}

In the following, we describe the procedures for generating the images from the hydrodynamic simulation used to train, validate, and test CNN. Figure~\ref{fig:making_images} illustrates the procedure described in the following.
\begin{itemize}
  \item{}[STEP 1] We define a $512^3$ grid in the simulation box. Then, we use the SPH kernel to calculate the HI number density $n_{\rm HI}$ and compute the differential brightness temperature $\delta T_b$, following our previous work \citep{ourwork} (see section~2.2 and 2.4), as
  \begin{equation} \label{eq:dTb}
    \delta T_b \sim \frac{3}{32 \pi} \frac{h_p c^3 A_{10}}{k_{\rm B} \nu^2_0} \left( 1 - \frac{T_\gamma (z)}{T_{\rm S}} \right) \frac{n_{\rm HI}}{(1+z)H(z)},
  \end{equation}
  where $\nu_0$ is the rest frequency of the 21cm signal, $T_\gamma$ is the temperature of CMB photon, $H(z)$ is the Hubble parameter, and $T_{\rm S}$ is the spin temperature of HI gas. In this calculation, we consider the redshift space distortion for the position of the simulation particle. The position is shifted along the line of sight (LOS) by $(1+z)v_{\parallel} / H(z)$, where $v_{\parallel}$ is the velocity of the simulation particle along the LOS.
    
  \item{}[STEP 2] 
  We have three choices for the LOS direction; these can be considered independent realizations. 
  We use the images generated for the two directions of LOS as the training and validation data, respectively and the rest of the LOS direction is used for the test data.
  We divide the simulation box along the LOS and generate the slices for each LOS direction. The width of each slice is $25.0$ $h^{-1}$ Mpc, corresponding to 2 slices. As the other choice of the width, we consider 3.125, 6.25, and 12.5 $h^{-1}$ Mpc, corresponding to the 16, 8, and 4 slices, respectively. We find the case of 25 $h^{-1}$Mpc projection shows the base performance. 
  
  For the data augmentation, we employ the offsets when we divide the simulation box
  to make the training or validation data. The offsets are
  $\Delta = 50 i / 512$ $h^{-1}$Mpc where $i$ is the integer from zero to $512/ \mathrm{(number\ of\ slices)}$ in the direction along the LOS. At the edge of the box, we apply the periodic boundary condition.
  This may increase the number of available images sufficiently and significantly help our training process to converge,
  although the shifted images are not independent of each other.
    
  \item{}[STEP 3] 
  $\delta T_b$ is projected onto the plane perpendicular to the LOS by summing up $\delta T_b$ along the LOS in each slice, i.e., $\delta T_b (\mbf{n}) = \sum_{\rm LOS} \delta T_b (\mbf{x})$.
  And then, in each slice, we cut out 8$\times$8 images. Therefore, the single image has $64^2$ pixels, 6.25\,$h^{-1}$Mpc on a side. Finally, we also apply the transformation of the pixel value following our previous work \citep{ourwork}, and it is described as $m_T (\mbf{n}) = \sinh^{-1} (\delta T_b(\mbf{n}) \ [\rm nK])$. Figure~\ref{fig:eg_images} shows examples of the image for the CDM, 2 keV FD, and 3 keV FD model at $z=3$ and $z=6$.

\end{itemize}

In total, we have 
$(8 \times 8)$ (cut out in STEP 3) $\times 512$ (offsets $\times$ slices) $\times 2$ (directions of the LOS) = 65535 for a set of training and validation data for one realization of the simulation. We use one percent of these images as the validation data. For the test data, we have $(8 \times 8)$ (cut out) $\times 1$ (direction of the LOS) = 64 images for a realization, where we do not apply the offsets. In addition, in training the CNN, 
the images are rotated every 90 degrees and flipped horizontally to generate another different set of images to increase the training images effectively.
\revi{In most cases, data is split into 80\% training and 20\% validation sets. However, due to our limited dataset, we chose an extreme split of 99\% for training and 1\% for validation. This decision aims to optimize the model as much as possible. While this approach may risk overfitting the training data, the independent hold-out test dataset, which is large enough, indicates that the model fits well and is not overfitted.}

\subsection{\label{ssec:noise}Noise budget}

\begin{figure*}
  \centering
  \includegraphics[keepaspectratio,width=\linewidth]{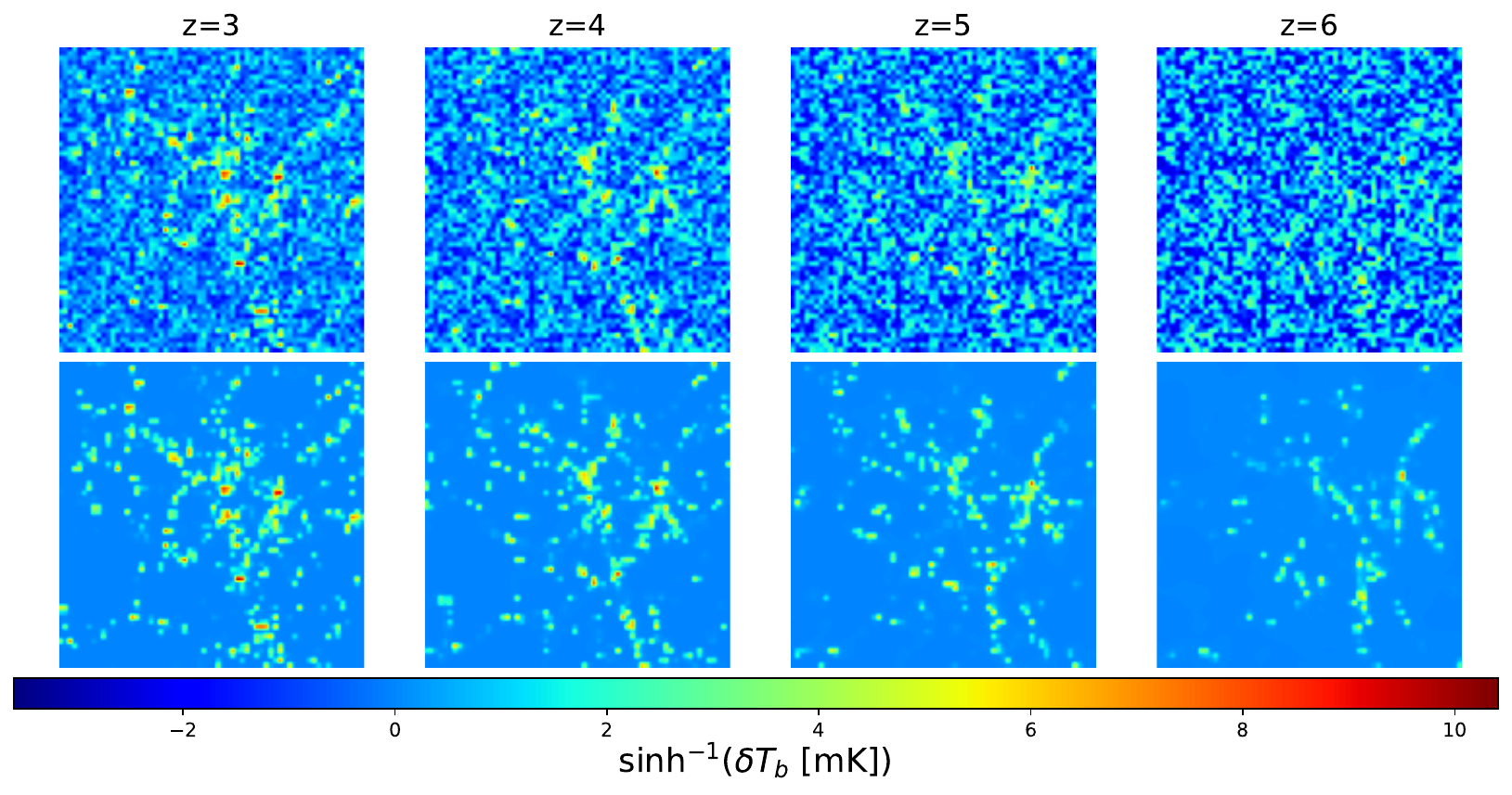}
  \caption{
  Examples of the noised image (top) and the signal-only image (bottom) in the same region for the CDM model. From left to right, each panel corresponds to $z=3,4,5,$ and 6. The size of the image corresponds to $6.25 \times 6.25\ [(h^{-1}Mpc)^2]$.
  }
  \label{fig:noised_images}
\end{figure*}

In this section, we add the system noise assuming SKA-Low to the images generated in Sec.~\ref{ssec:images} to test the effect of the noise on the CNN classification between the CDM and FD model. 

\revi{Here we assume }the noise at a pixel follows the Gaussian, where its mean is zero, and variance is given by \citep{2004ApJ...608..622Z}
\begin{equation}
\label{eq:pixel_noise}
\sigma^2_N = d^2 u C^{N}_{l}.
\end{equation}
$du$ corresponds to the angular size of the simulation box in Fourier space and is $2\pi \chi / L_{\mathrm{box}}$, where $\chi$ and $L_{\mathrm{box}}$ are the comoving distance to the source of the signal and the box size of our simulation, respectively.

$C^{N}_{l}$ is the angular noise power spectrum. In this work, we simply assume that $C^{N}_{l}$ is scale independent and is described as \citep{2016MNRAS.459..863P}
\begin{equation}
C^{N}_{l} = \frac{T^{2}_{\mathrm{sys}} [\mathrm{FOV}]^2}{B t_0 n(u)},
\end{equation}
where $B$ is the frequency bandwidth corresponding to the projection length in our work, and $t_0$ is the integration time. In this work, we assume $t_0=1000$ hours.
FOV is the field of view and is approximated as $\lambda^2 / A_{\rm d}$, where $\lambda$ is the observed wavelength and $D_{\rm d}$ is the diameter of the single antenna.

$T_{\mathrm{sys}}$ is the system temperature and is described by using the observed frequency $\nu = \nu_0 / (1+z)$ as
\begin{equation}
    T_{\mathrm{sys}} = 28 + 66 \left( \frac{\nu}{300\ \mathrm{MHz}} \right)^{-2.55}\ \mathrm{K}.
\end{equation}
$n(u)$ is the number density of the baselines, described as \citep{2021MNRAS.500.3162B}
\begin{equation}
    n(u) = \frac{N_{\mathrm{d}} (N_{\mathrm{d}} - 1)}{2 \pi (u^2_{\mathrm{max}} - u^2_{\mathrm{min}})},
\end{equation}
where $N_{\rm d}$ is the number of the antenna, $u_{\rm max}$ and $u_{\rm min}$ are defined by using the maximum baseline $D_{\rm max}$ and the antenna diameter $D_{\rm d}$ as $D_{\rm max} / \lambda$ and $D_{\rm d} / \lambda$, respectively.

Our work considers the redshifts from $z=6$ to $z=3$ with the SKA-Low. To compute Eq.~(\ref{eq:pixel_noise}), we approximate $A_d \sim A_{\rm tot} / N_{\rm d}$, where $A_{\rm tot}$ is the total collective area and $A_{\rm d} = \pi \left( D_{\rm d} / 2 \right)^2$. We use $A_{\rm tot}=419000\ \mathrm{m}^2$, $N_{\rm d} = 512$, and $D_{\rm max} = 74$ km as the SKA-Low configuration \footnote{https://www.skao.int/en/explore/telescopes/ska-low}.

To generate noised images, we make the noise-only image with $512 \times 512$ pixels, corresponding to $50 \times 50 \ (h^{-1}\mathrm{Mpc})^2$, and add this image to the $\delta T_b$ map after the projection in the STEP 3 in Sec.~\ref{ssec:images}. In addition, we apply the transformation by $m^{\prime}_T (\mbf{n}) = \sinh^{-1} (\delta T_b(\mbf{n}) \ [\rm mK])$ instead of $m_T (\mbf{n})$ in STEP 3, 
because we found that our CNN shows better performance for the classification of the noised images when we apply $m^{\prime}_T (\mbf{n})$.

Figure~\ref{fig:noised_images} shows examples of noised images. We can see the noise at the higher redshift hides the signals because the structure has not grown yet compared to the lower redshift.

%

%
\section{\label{sec:method}Methods}
%

This work considers the $N_{\rm DM}$-class classification, where $N_{\rm DM}$ is the number of the dark matter models assumed in the classification. Specifically, the 2-class (binary) classification of the CDM and FD WDM (Sec. \ref{ssec:bn_dTb_image} and \ref{ssec:bn_dTb_image_w/noise}) and the 5-class classification of the CDM, FD, $\nu$MSM, SF, and DW models (Sec. \ref{ssec:clasify_wdm_models}).
In the section~\ref{ssec:pk} and \ref{ssec:CNN}, we consider the analysis focusing on images and power spectra from a single redshift, which is one of $z = [3, 4, 5, 6]$. And section~\ref{ssec:joint-z} considers that all images at $z=[3,4,5, 6]$ are simultaneously used as the input to the machine learning algorithms. Finally, Sec.~\ref{ssec:training} shows the training procedure.

\subsection{\label{ssec:pk} Power spectrum}

\begin{figure}
    \centering
    \includegraphics[width=\linewidth]{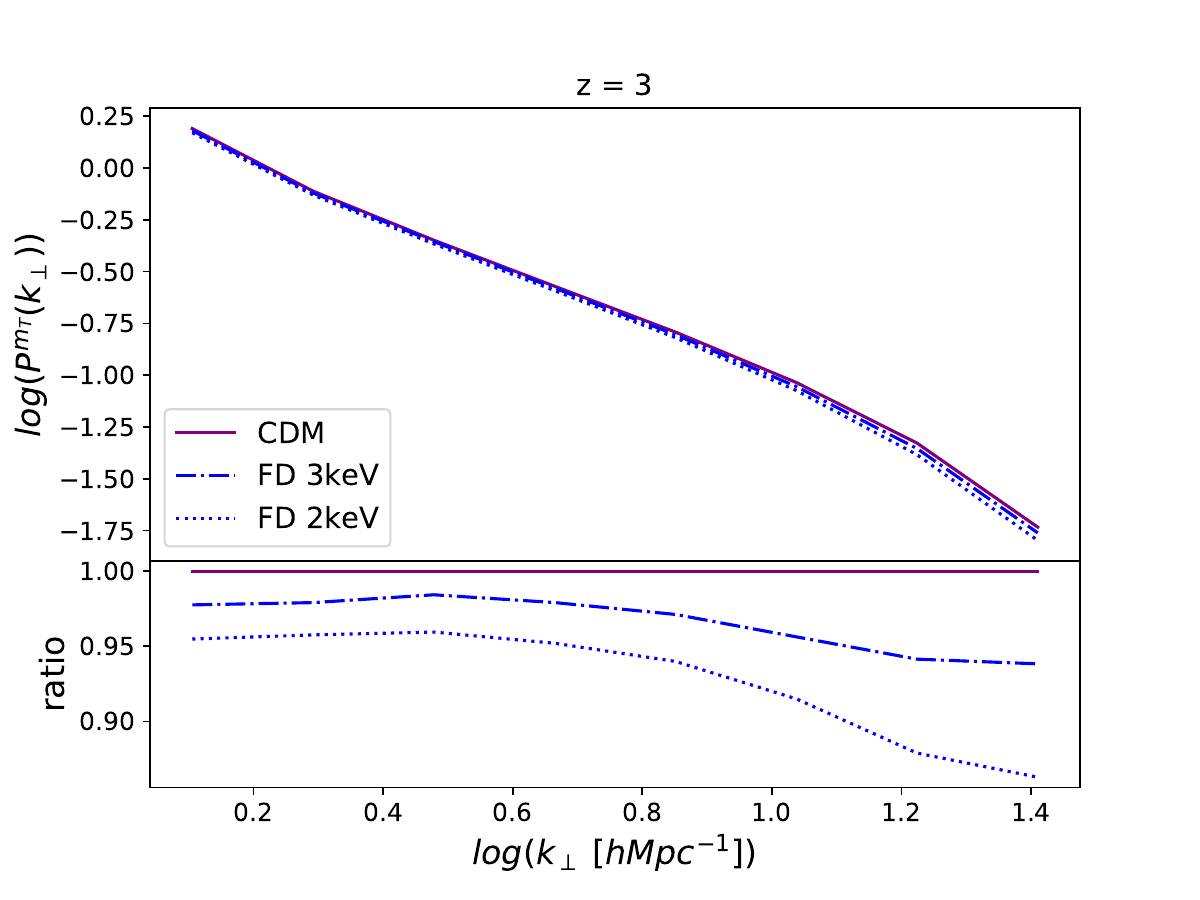}
    \caption{
    The upper panel shows the $m_T$ power spectra of images for the CDM, 3keV FD, and 2keV FD model at $z=3$. In the lower panel, we show the power spectrum ratio for FD models to that for the CDM model.
    }
    \label{fig:pk_dTb}
\end{figure}

\begin{figure}
    \centering
    \includegraphics[width=\linewidth]{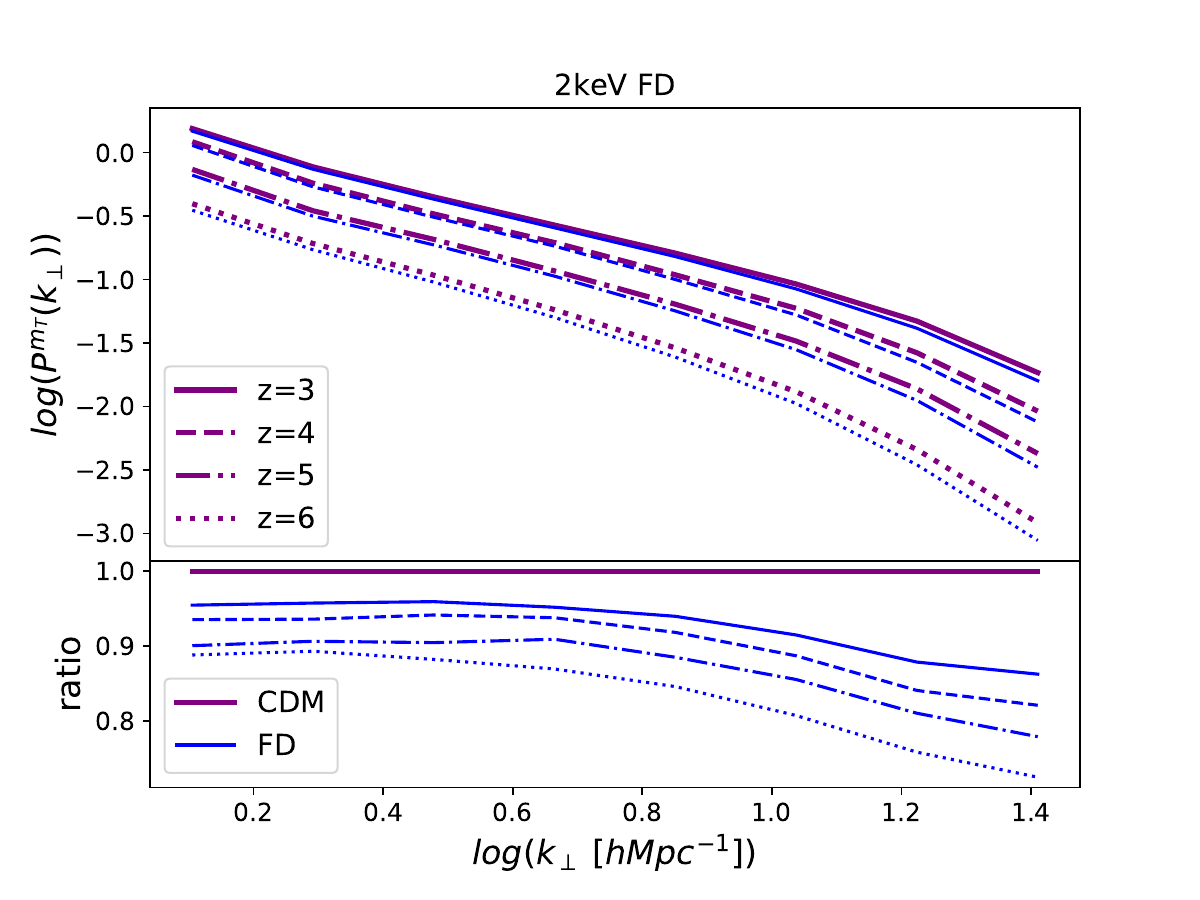}
    \caption{
    The upper panel shows the $m_T$ power spectra of images for the CDM (purple) and 2keV FD model (blue) at $z=$3 (solid), 4 (dashed), 5 (dot-dashed), and 6 (dotted). In the lower panel, we show the power spectrum ratio for FD models to that for the CDM model for each redshift.
    }
    \label{fig:pk_dTb_z}
\end{figure}

To compare the image-based analysis with the power spectrum analysis, we implement the neural network trained by the power spectrum. In this section, we show the calculation of the power spectra of images and the architecture of our neural network.

To calculate the power spectra of images, we apply a two-dimensional Fourier transform to the images, and the pixel values of the Fourier-transformed image are calculated as 
\begin{equation}
       \tilde{m}_T (\mbf{k}_\perp) = \int \exp{(-i \mbf{k} \cdot \mbf{n})} m_T (\mbf{n}) d^2 n,
\end{equation}
where $k_\perp$ is the wave number perpendicular to the LOS.

And then, the power spectrum of a $m_T (\mbf{n})$ image is written as 
\begin{equation} \label{eq:pk_dTb}
 P^{m_T}(k_i) = \frac{1}{L^2} \frac{1}{N
_{k_i}} \sum_j \tilde{m}_T (\mbf{k}_j) \tilde{m}_T^*(\mbf{k}_j),
\end{equation}
where $k_i = |\mbf{k}_i|$ is the absolute value of the wave number of the center of the $i$-th bin, $\mbf{k}_j$ is the wave number satisfies $k_i \le |\mbf{k}_j| < k_{i+1}$, $N_{k_{i}}$ is the number of the modes in $i$-th $k$ bin and $L$ is the size of image and is 6.25 $h^{-1}$Mpc. 
The minimum and maximum wave numbers are determined by the size and resolution of the images from which we measure the spectrum, and they are $k_{\rm min}=2\pi/6.25\ h \mathrm{Mpc}^{-1}$ and  $k_{\rm max}=2\pi/(6.25/64)\ h \mathrm{Mpc}^{-1}$, respectively.
We change the number of $k$-bins from 2 to 16 and find that 8 minimizes the AUC value. Therefore, we apply 8 bins.

\begin{table}[h]
  \begin{tabular}{| c || c | c |}
    \hline
    \ & Layer & Output size \\ \hline
    1 & Input & 8 \\
    2 & fully connected & 64 \\
      & \revi{dropout(0.4)} & \\
    3 & fully connected & 128 \\
      & \revi{dropout(0.4)} & \\
    4 & fully connected & 256 \\
      & \revi{dropout(0.4)} & \\
    5 & fully connected & 512 \\
      & \revi{dropout(0.4)} & \\
    6 & fully connected & $N_{\rm DM}$ \\ \hline
  \end{tabular}
  \caption{
    Our neural network architecture. This network has 174,530 trainable parameters. After the 2nd, 3rd, 4th, and 5th layers, we apply the dropout, and 40 \% of the nodes selected randomly in the previous layer are not used in training.
  }
  \label{tb:NN_arch}
\end{table}

Figure~\ref{fig:pk_dTb} shows the power spectra calculated by Eq.~(\ref{eq:pk_dTb}) for CDM and FD with some masses and the ratio of the FD power spectrum to the CDM's one. For the power spectrum of the HI distribution, we can see the suppression of the amplitude at all scales in images differently from the dark matter case shown in Figure~\ref{fig:pk_wdm_mass}. The relation between the power spectrum of the matter density fluctuation $\delta_{m}$ and the one of the differential brightness temperature $\delta T_b$ is 
\begin{align}
    \langle \delta \tilde{T}_b \delta \tilde{T}_b \rangle &\propto \langle \tilde{n}_{\rm HI} \tilde{n}_{\rm HI} \rangle \nonumber \\
    &= \langle (\bar{n}_{\rm HI} \tilde{\delta}_{\rm HI} + \bar{n}_{\rm HI}) (\bar{n}_{\rm HI} \tilde{\delta}_{\rm HI} + \bar{n}_{\rm HI}) \rangle \nonumber \\
    & = \bar{n}^2_{\rm HI} b^2_{\rm HI} \langle \tilde{\delta}_{m} \tilde{\delta}_{m} \rangle + \bar{n}^2_{\rm HI},
\end{align}
where $\tilde{A}$ represents the Fourier counterpart of a physical quantity $A$, $\bar{n}_{\mathrm{HI}}$ is the mean of the HI number density, and $b_{\rm HI}$ is the HI bias. Therefore, even when the suppression of the matter power spectrum caused by WDM appears only at the small scale, the overall amplitude of $\delta T_b$ power spectrum is affected due to the smaller HI number density in the WDM models. Figure~\ref{fig:pk_dTb_z} also shows the $P^{m_T} (k_\perp)$ for the CDM and 2 keV FD model at redshift $z=$3, 4, 5, and 6. We can see the growth of structure in the upper panel and the dependence of the power spectra on the redshift.

For the power spectrum analysis, we use a neural network. We implement the neural network with \texttt{PYTORCH} \citep{pytorch}, which is a \texttt{Python} module to implement the deep learning algorithm, shown in Table~\ref{tb:NN_arch}. This neural network has four hidden fully connected layers and we apply the 40 \% dropout \citep{srivastava2014dropout} after each layer to avoid overfitting to the training data. Our neural network is trained by the power spectrum of the training images generated in Sec.~\ref{ssec:images}. Therefore, we have exactly the same number of training, validation, and test datasets for the power spectrum as in the case of the CNN. We examine some architecture with more or less layers and nodes and find the architecture shown in Table.~\ref{tb:NN_arch} shows the best performance in the binary classification of the CDM model and 2keV FD model.

The last layer of our neural network has $N_{\rm DM}$ nodes. For the $i$-th input image of the model M, we describe the output from the $k$-th node as $y_{i}(k|{\rm M})$. Each node corresponds to the output for each dark matter model, and the weight parameters are optimized to maximize $y_i (\rm M|M)$ by training.

\subsection{\label{ssec:CNN} CNN}

\begin{table}[h]
  \begin{tabular}{| c || c | c |}
    \hline
    \ & Layer & Output map size \\ \hline
    1 & Input & $64 \times 64 \times 1$ \\
    2 & \revi{BN($3 \times 3$ conv.)} & $62 \times 62 \times 64$ \\
    3 & \revi{BN($3 \times 3$ conv.)} & $60 \times 60 \times 64$ \\
    4 & \revi{BN($3 \times 3$ conv.)} & $58 \times 58 \times 128$ \\
    5 & \revi{BN($3 \times 3$ conv.)} & $56 \times 56 \times 128$ \\
    6 & \revi{BN($3 \times 3$ conv.)} & $54 \times 54 \times 256$ \\
    7 & \revi{BN($1 \times 1$ conv.)} & $54 \times 54 \times 128$ \\
    8 & \revi{BN($3 \times 3$ conv.)} & $52 \times 52 \times 256$ \\
    9 & \revi{BN($3 \times 3$ conv.)} & $50 \times 50 \times 512$ \\
    10 & \revi{BN($1 \times 1$ conv.)} & $50 \times 50 \times 256$ \\
    11 & \revi{BN($3 \times 3$ conv.)} & $48 \times 48 \times 512$ \\
    12 & $2 \times 2$ AveragePooling & $24 \times 24 \times 512$ \\
    13 & GlobalAveragePooling & 1 $\times$ 1 $\times$ 512 \\
    14 & fully connected & $N_{\rm DM}$ \\ \hline
  \end{tabular}
  \caption{
  Our CNN architecture. The output map size represents the $\mathrm{(height)} \times \mathrm{(width)} \times \mathrm{channels)}$ of the output from the layer, where channel means the number of the kernels. After each convolution layer, we apply batch normalization \revi{(BN)}. This CNN has the 3,379,522 trainable parameters.
  }
  \label{tb:CNN_arch}
\end{table}

In this work, we use CNN for the image-based analysis. 
Table~\ref{tb:CNN_arch} shows the architecture of our CNN\revi{, which was originally proposed by the convergence field analysis \citep{2019MNRAS.490.1843R}.} We explore some deeper or shallower architecture. In the following, we apply the architecture shown in Table~\ref{tb:CNN_arch}, which shows the best performance among our tests for the binary classification of the CDM and 2keV FD WDM. 
\revi{In addition, the choices of the kernel size or the channel number are also subject to optimization. We find that the choice of kernel size and number of channels mildly affects the results, but the current choice is the most effective regarding the DM model classification.}

Our CNN consists of 10 convolution layers, an AveragePooling layer, and 
a GlobalAveragePooling layer \citep{2013arXiv1312.4400L}. In the $x \times y$ convolution layer, the input to this layer is convolved by the $x \times y$ kernel with the stride $=1$. In our CNN, the padding is not applied, and the image size becomes smaller after each $3 \times 3$ convolution layer. Additionally, we apply the batch normalization \citep{BatchNorm} to improve the efficiency of the training.

This CNN has $\sim 3.4 \times 10^6$ trainable parameter, and we have $6.5 \times 10^4$ for each dark matter model. Therefore, we train our CNN by $\sim 10^5$ training images for the binary classification and have enough training data \citep{Han2015}.

The output of the last layer represents the same as our neural network in Sec.~\ref{ssec:pk}, and we also describe it as $y_{i}(k|{\rm M})$.

\subsection{\label{ssec:joint-z} Joint redshift}

\begin{figure*}
  \centering
  \includegraphics[keepaspectratio,width=\linewidth]{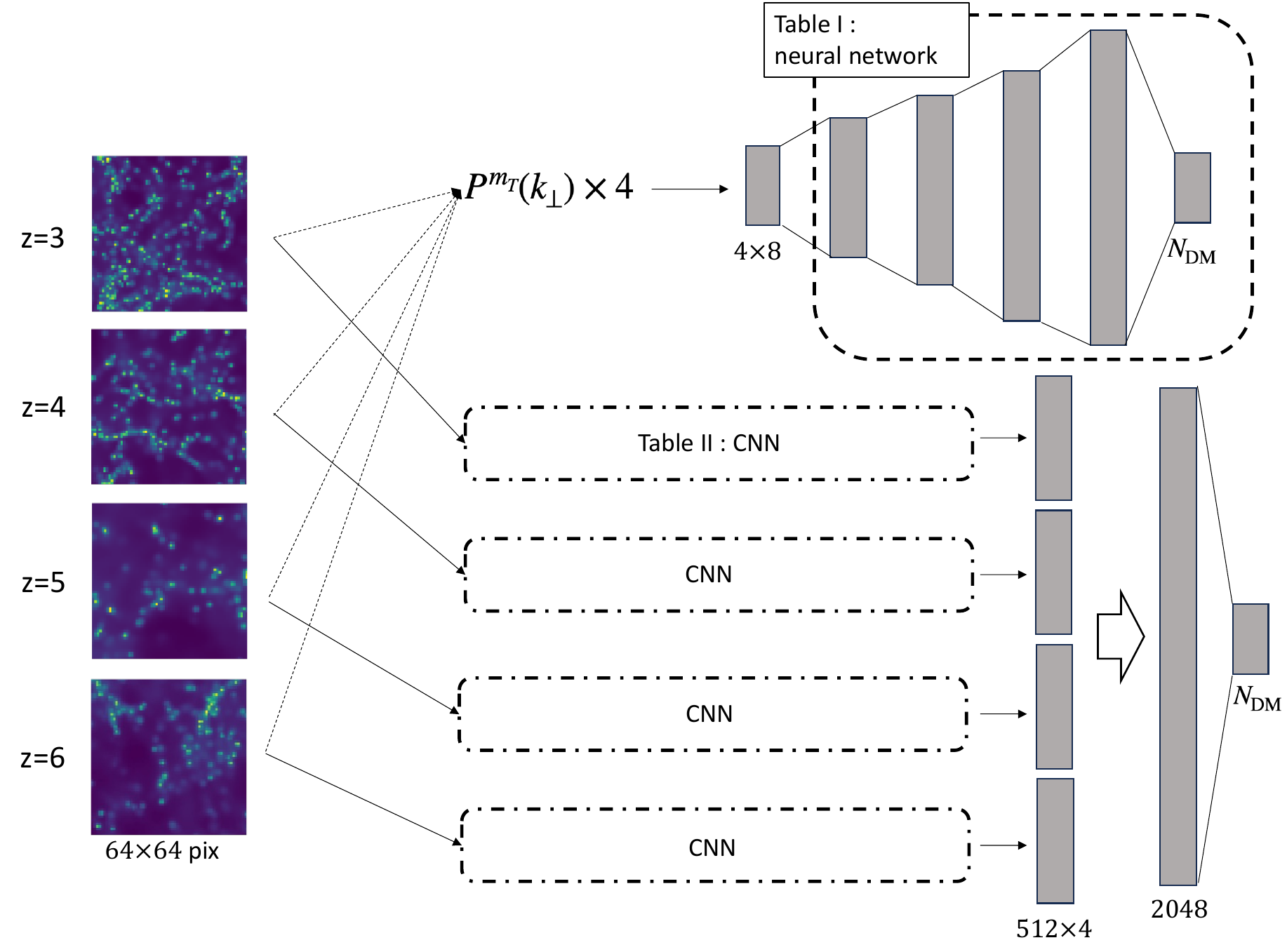}
  \caption{
  The architectures for joint redshift analysis are shown. For the power spectrum analysis, the power spectra from each redshift are concatenated, and we input it to the neural network shown in Table~\ref{tb:NN_arch}. For the image-based analysis, we construct 4 CNNs of Table~\ref{tb:CNN_arch} and input the image from each redshift to each CNN. And then, the outputs from the CNNs are concatenated by a fully connected layer.
  }
  \label{fig:joint-z_arch}
\end{figure*}

Here, we investigate the joint redshift analysis for both the power spectrum analysis with the neural network and the image-based analysis with CNN. By using the power spectra and images at different redshifts $z=[3,4,5,6]$ simultaneously as the input to the neural network and CNN, we examine whether the classification results are improved due to the increase of information included in the inputs. In the following subsection, we randomly pick up four images from each redshift and use them as input for our machine-learning algorithms.

For the power spectrum analysis with the neural network, we first calculate $P^{m_T} (k_\perp)$ for the input images at each redshift. Then, these power spectra are simply concatenated and used as the input to the neural network. The architecture is the same as the one shown in Table~\ref{tb:NN_arch} except for the input layer, and the input size is 8 (number of $k_\perp$ at a single redshift) $\times$ 4 (number of redshifts) $=32$. We examine the architectures including more nodes than Table~\ref{tb:NN_arch}, but we do not find an improvement in the classification result.

For the image-based analysis with CNN, we prepare 4 CNNs, whose architecture is the same as Table~\ref{tb:CNN_arch} except that it does not have a last fully connected layer. Subsequently, each CNN is fed with images from each redshift, and their outputs are combined by an additional fully connected layer, which has 512 (number of feature maps 
 from GlobalAveragePooling layer) $\times$ 4 (number of redshifts) $= 2048$ nodes.

 The architectures for the joint redshifts analysis are shown in Figure~\ref{fig:joint-z_arch}.

\subsection{\label{ssec:training} Training}

For the evaluation of the output from the neural network and CNN, we convert the outputs 
by the softmax function as
\begin{equation} \label{eq:softmax}
  p_i(k|\mathrm{M}) = \frac{\exp{(y_i(k|\mathrm{M}))}}{\sum_l \exp{(y_{i}(l|\mathrm{M}))}},
\end{equation}
where we sum up the outputs over all nodes in the last layer in the denominator, and $p_{i}(k|\rm M)$ is the probability predicted by our CNN that the $i$-th input image is $k$ model and M means the true dark matter model for the $i$-th input.

To optimize our neural network and CNN, we update the weight parameters in our machine learning architectures to maximize the predicted probability for the correct model M for the $i$-th input image $p_{i} (M|M)$. To do this, we minimize the loss function.
In this work, we adopt a typical cross-entropy as a loss function,
\begin{equation}
  \label{eq:loss}
  E_i(\boldsymbol{w}) = - \sum_{k} \tilde{p}_i({k|{\rm M}}) \ln{(p_i(k|{\rm M}))}.
\end{equation}
In this equation,
$\tilde{p}_i$ takes 1 for correct class ($k={\rm M}$) and 0 otherwise ($k\neq{\rm M}$), and predicted probability $p_i$ takes continuous values between 0 and 1. The output $p_i$ is an implicit function of the weight parameters $\boldsymbol{w}$.

For the optimization, we apply the AMSG{\footnotesize RAD}\cite{AMSGRAD} and take the learning rate as $10^{-4}$ and $10^{-5}$ for the neural network and CNN, respectively. In training, we apply the mini-batch learning with the batch size $= 32$.

%
\section{\label{sec:result}Results}
%

In this section, we show the results of the dark matter model 
classifications. First, we show the results of the binary ($N_{\rm DM} = 2$) classification between the CDM and FD WDM in Sec.~\ref{ssec:bn_dTb_image}. Sec.~\ref{ssec:clasify_wdm_models} discusses the classification including different WDM models. 

\subsection{\label{ssec:bn_dTb_image} Binary classification}

\begin{figure*}
\centering
\begin{tabular}{cc}
  \includegraphics[keepaspectratio,width=0.5\linewidth]{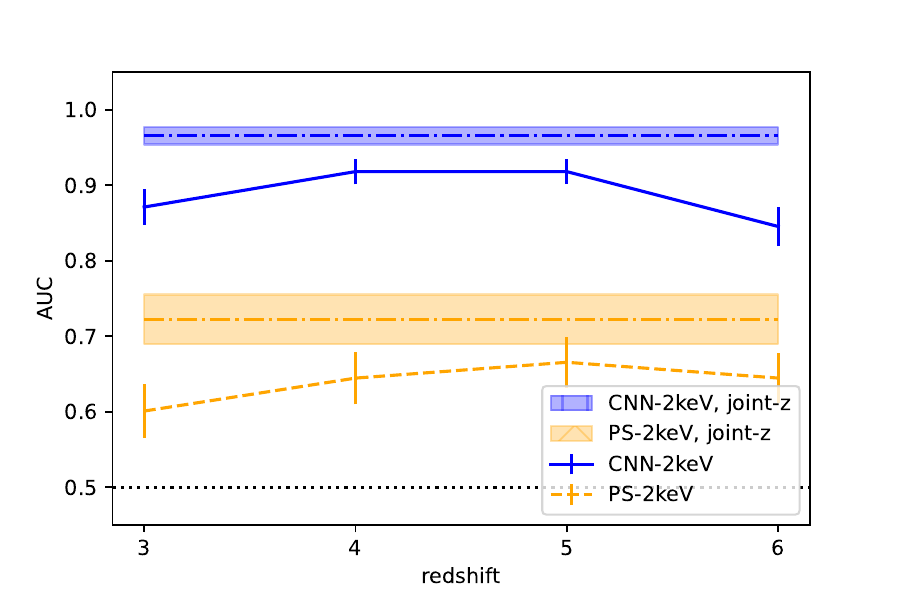} &
  \includegraphics[keepaspectratio,width=0.5\linewidth]{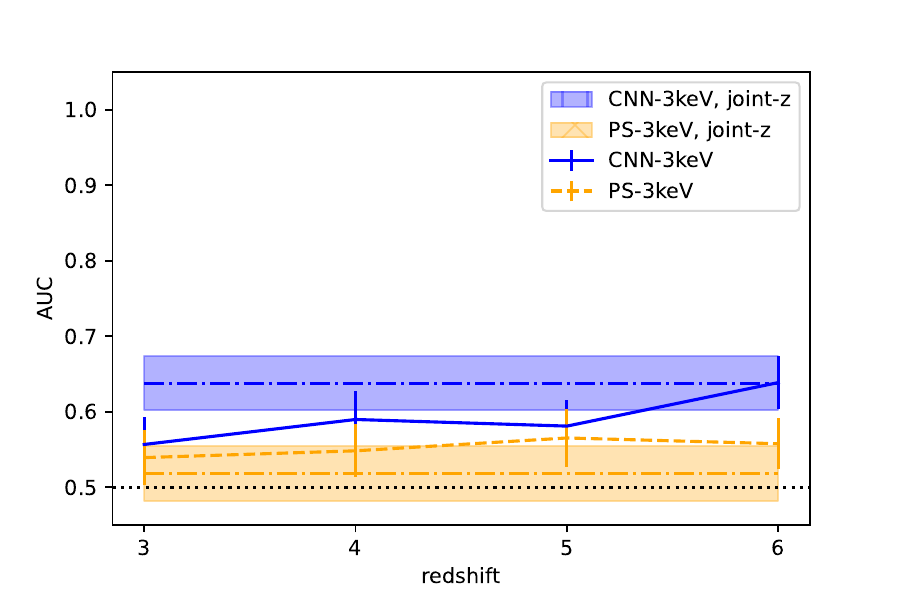}
\end{tabular}
  \caption{
  The AUC values for the classifications of the test images between the CDM and FD model with CNN (blue) and the power spectrum (orange) at each redshift. The left and right panels correspond to the classification for the 2 keV and 3 keV FD models, respectively. In addition, the dot-dashed line and shaded region represent the joint redshift analysis. The error bar and shaded region show the 1$\sigma$ jackknife error.
  } 
  \label{fig:AUC_dTb}
\end{figure*}

Here, we classify the CDM and FD WDM images and assume the FD WDM mass is 2 and 3 keV.

We use the neural network and CNN to classify the images and evaluate the results using the area under the curve (AUC) of the receiver operating characteristic (ROC) curve. In calculating the AUC following \citep{ourwork}, we assume that the positive and negative represent the input image is classified into the FD and CDM model, respectively. If our neural network or CNN can accurately classify the images, AUC becomes close to 1. On the other hand, AUC is around 0.5 for the binary classification when they cannot classify the images.

Figure~\ref{fig:AUC_dTb} shows the AUC values for the binary classification with CNN and power spectrum denoted as PS. The left and right panels correspond to the classification for 2 keV and 3 keV FD WDM, respectively. In this figure, the errors (1-$\sigma$) are estimated by the jackknife resampling of the test images, where the number of resampled images is 127 for each dark matter model. 

For both 2\,keV and 3\,keV FD WDM masses, CNN (blue) shows larger AUC values than the power spectrum (orange) at each redshift. Our CNN can classify the CDM and 2\,keV FD WDM with AUC $\sim 0.9$, while the AUC for the power spectrum is $\sim 0.6$ at all redshifts we consider. This is not surprising because CNN uses images that can include information beyond the power spectrum, such as higher-order statistics.
For the 3\,keV FD case, CNN shows a slightly larger AUC than the power spectrum, and it is difficult to distinguish 3\,keV FD WDM from CDM. \revi{It should be clear from Figure \ref{fig:pk_dTb} that the power spectrum of $m_T$ for 3keV differs from the CDM model at most 5\% and the most significant difference is only on smaller scales where the shot noise dominates. Therefore, it is more challenging to discriminate the 3keV model from the CDM compared to the 2keV model.}

Additionally, while Figure~\ref{fig:pk_dTb_z} shows the redshift dependence of the ratio of the power spectrum of the CDM and FD WDM, we do not find significant effects on our results for the classifications. The halo number is smaller at higher redshifts, and the shot noise in the images is more significant. Therefore, the classification results are not improved despite the larger difference in the power spectrum at higher redshifts.

We also show the results for the joint redshift analysis with the shaded region in Figure~\ref{fig:AUC_dTb}. For the classification between the CDM and 2keV FD WDM, we find the AUC for the joint redshift is larger than the one for the single redshift at each redshift and the improvement of the AUC for both CNN and the power spectrum. 
On the other hand, the power spectrum information alone is insufficient to classify the CDM and 3 keV FD model for the 3 keV FD WDM, and AUC does not improve even with joint redshift analysis.

\subsection{\label{ssec:bn_dTb_image_w/noise} Binary classification with the system noise}

\begin{figure*}
\centering
\begin{tabular}{cc}
  \includegraphics[keepaspectratio,width=0.5\linewidth]{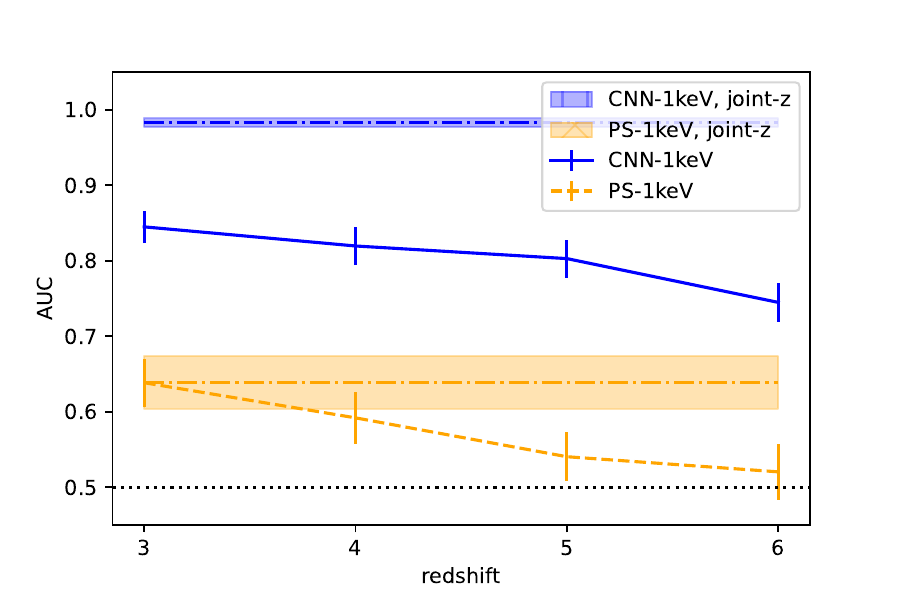} &
  \includegraphics[keepaspectratio,width=0.5\linewidth]{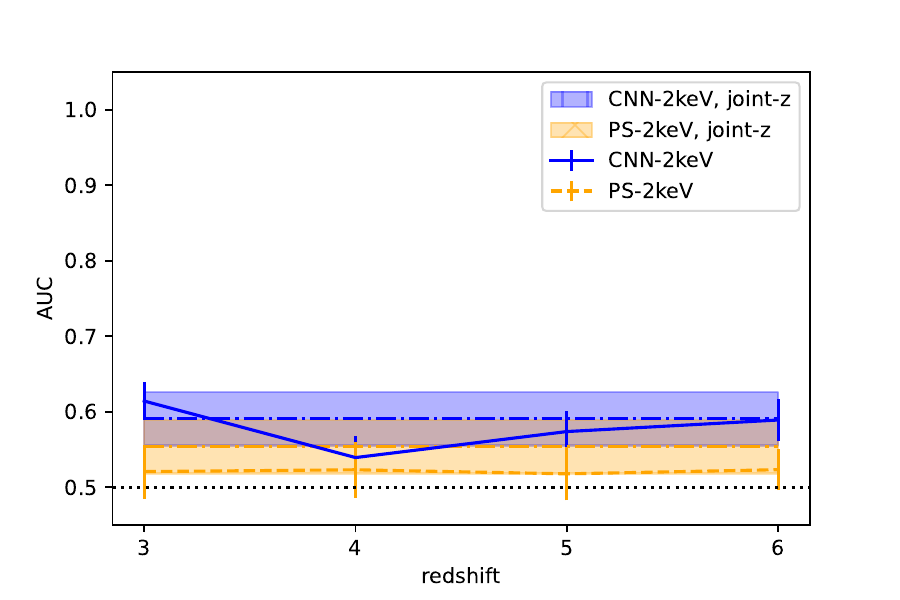}
\end{tabular}
  \caption{
  The AUC values for the classification of the test images between the CDM and FD model with CNN (blue) and the power spectrum (orange) at each redshift. The left and right panels correspond to the classification for the 1 keV and 2 keV FD models, respectively. In addition, the dot-dashed line and shaded region represent the joint redshift analysis. The error bar and shaded region show the 1$\sigma$ jackknife error.
  } 
  \label{fig:AUC_dTb_noise}
\end{figure*}

Here, we consider the binary classification between the CDM and FD WDM model for the noised images to compare with the previous section without noises. 

Figure~\ref{fig:AUC_dTb_noise} shows the AUC values for the classification between the CDM and FD model. The left and right panels correspond to the 1 keV and 2 keV FD WDM masses.

For 1 keV mass, the AUC for the power spectrum depends on the redshift. The AUC is 0.64 and the power spectrum can classify the images at $z=3$, while the AUC is 0.52 at $z=6$. These results are consistent with Figure~\ref{fig:noised_images}, where the signals are hidden by the noise at higher redshift. 
The joint redshift analysis shows a similar AUC value to the single redshift analysis at $z=3$, and the other redshift does not contribute to the classification.

CNN shows a better performance than the power spectrum analysis for 1 keV FD WDM mass. At z=$3$, the AUC for CNN is 0.84, while the one for the power spectrum is 0.64. CNN's results also show the redshift dependence, but the dependence is not as strong as the case of the power spectrum. This is not surprising because CNN can extract information beyond the power spectrum.  For instance, the CNN can extract the shape information such as the Minkowski functionals as demonstrated in Reference \cite{2020PhRvD.102l3506M}.

The joint redshift analysis shows better AUC than the single redshift analysis. The AUC value is 0.98, and the CNN can accurately distinguish the images. 

For the 2 keV FD WDM mass, the AUC for the power spectrum is not significantly larger than 0.5 at each redshift, and the power spectrum cannot classify the images. Even for CNN, the AUC is slightly larger than 0.5, and the classification would be difficult with the images including the noises. 

\subsection{\label{ssec:clasify_wdm_models} Classification of WDM models}

\begin{figure*}[ht]
\begin{center}
    \includegraphics[width=\linewidth]{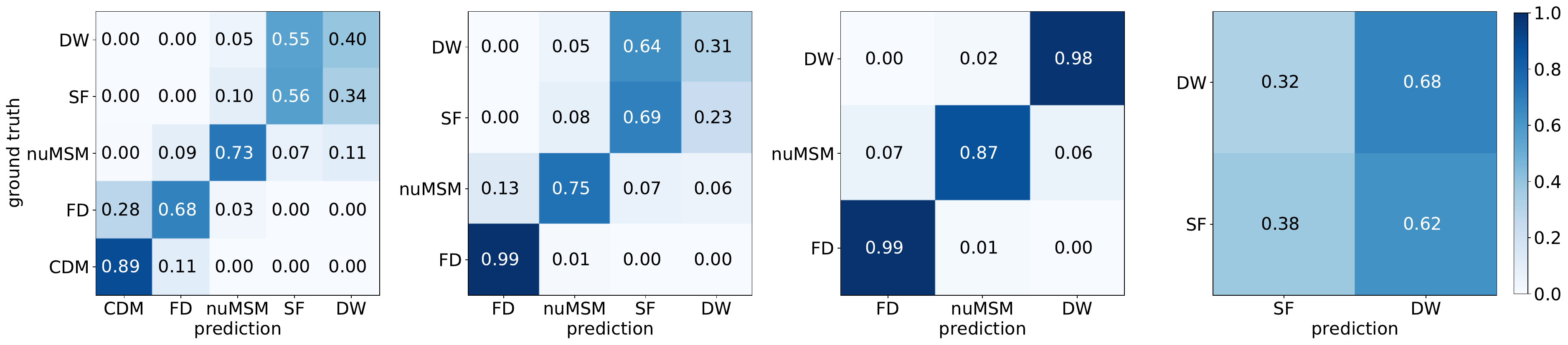}
\end{center}
\caption{
This figure shows the $\bar{p} (k|\rm M)$ for each input dark matter model for $m_{\rm DM}=2$\,keV. The 5-, 4-, 3-, and 2-class classifications correspond to the \revi{left to} right panel, respectively. In each panel, the vertical axis shows the ground truth model of the input images, and the \revi{horizontal axis represents the prediction of the model}.
}
\label{fig:wdm_classification_2keV}
\end{figure*}

\begin{figure*}[ht]
\begin{center}
        \includegraphics[width=\linewidth]{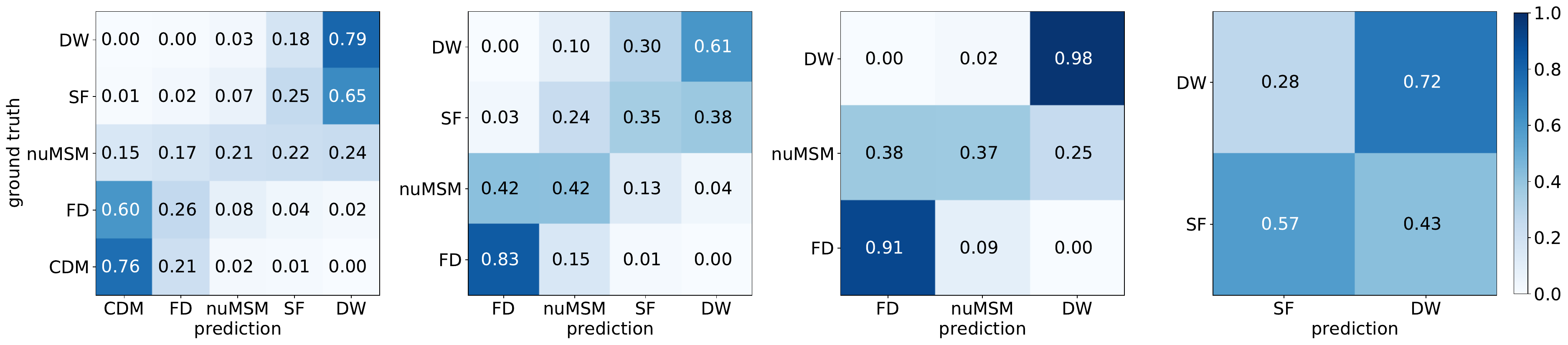}
\end{center}
\caption{
Same as Figure~\ref{fig:wdm_classification_2keV} but for $m_{\rm DM} =3$ keV.
}
\label{fig:wdm_classification_3keV}
\end{figure*}

Here, we consider the classification of the dark matter models for $N_{\rm DM} \ge 2$. 
We additionally conduct the simulations for $\nu$MSM, DW, and SF models introduced in Sec.~\ref{sec:introduction} and study the cases of 2 and 3 keV WDM masses. The configuration of the simulations is the same as the one in Sec.~\ref{ssec:simulation}. 
The initial power spectra for each WDM model are computed by the same procedure for the FD model as shown in Sec.~\ref{sec:other_wdm_models}.

In this section, we classify the images with the joint redshift analysis and, for simplicity, do not consider the system noise. To evaluate the results, we consider the mean probability of the results as
\begin{equation}
    \bar{p} (k | \mathrm{M}) = \frac{1}{N_{\mathrm{M}}} \sum_i p_i (k|\rm M),
\end{equation}
where $N_{\rm M}$ is the number of the test images of the dark matter model M. $\bar{p} (k | \mathrm{M})$ is maximized when $k=\mathrm{M}$.

In the following, we consider the 4 cases: 5-class (CDM + 4 WDM), 4-class (4 WDM only), 3-class (FD, $\nu$MSM, and DW), and 2-class (SF and DW). Here, we fix the WDM particle mass $m_{\rm DM}$, corresponding to $m_{\rm FD}$ and the left-hand side of Eq.~(\ref{eq:FD-DW})-(\ref{eq:FD-nuMSM}) for each WDM model, and we calculated the initial power spectrum of our simulation following Eq.~(\ref{eq:ratio_wc})-(\ref{eq:Phi-for-Transfer}).

As we can see in Figure~\ref{fig:k_half}, which shows the relation between $m_{\rm DM}$ and $k_{1/2}$ following Eq.~(\ref{eq:k_half}), the lighter dark matter can be distinguished more easily from the CDM model due to the larger scale of the free streaming. On the other hand, the difference between the WDM models becomes larger in the case of the heavier WDM. Especially, SF and DW are similar at around $m\sim 2$ keV. Therefore, we examine the case of the 3-class classification, which does not classify the SF and DW, and the binary classification between the SF and DW model.

Figure~\ref{fig:wdm_classification_2keV} shows the results 
of discriminating the different model for given $m_{\rm DM}=2 {\rm keV}.$ 
The \revi{left} panel corresponds to the 5-class classification. The CDM, FD, and $\nu$MSM models can be distinguished from the other models, while the DW model cannot be distinguished from the SF model. In the 2-class classification between the SF and DW model \revi{(right panel), the CNN predicts the model with similar probability, regardless of which model the image is generated from, which means it cannot classify these two models correctly.}
As we can see in Figure~\ref{fig:k_half}, the SF and DW models have similar features for $m_{\rm DM}=$ 2 keV because their initial power spectrum is almost identical; hence, it is difficult to distinguish these models even for the image-based analysis. 

In Figure~\ref{fig:wdm_classification_2keV}, the \revi{second panel from the right} shows the 4-class classification among WDM models. Our CNN can maximize $\bar{p} (k|\mathrm{M})$ for the correct model for the FD and $\nu$MSM model, and the FD model is distinguished from the other models with higher accuracy because our CNN is not perplexed by the CDM model. Finally, to illustrate the degeneracy between the SF and DW models, we also showed the 3-class classification results between the FD, $\nu$MSM, and DW model. As a result, our CNN can distinguish the FD, $\nu$MSM, and the other models (SF and DW), while it is difficult to distinguish the SF and DW models.

Figure~\ref{fig:wdm_classification_3keV} shows those same as Figure~\ref{fig:wdm_classification_2keV}, but we assume $m_{\rm DM}=3$ keV. For this case, the difference between the dark matter models appears at the smaller scales than the case of $m_{\rm DM}=2$ keV. For the 5-class classification (top left), our CNN cannot distinguish the FD from the CDM model as we can see in the second raw corresponding to the case that the input is the FD model.
The free streaming scale is smaller than the 2 keV case and the classification between the CDM and FD becomes difficult. The $\nu$MSM cannot be distinguished either from the other models, and the $\bar{p} (k|\nu \mathrm{MSM})$ is a similar value regardless of the target model represented by $k$. The amplitude of the power spectrum for the $\nu$MSM model is in between the other dark matter models, and this model cannot be distinguished from the others. In the 4-class classification, the FD model can be distinguished from the other models, but $\bar{p} (\mathrm{FD}|\nu\mathrm{MSM})$ shows a comparable value to $\bar{p} (\nu\mathrm{MSM}|\nu\mathrm{MSM})$. The SF and DW models cannot be distinguished and are confused by the $\nu$MSM model.

For the 2-class classification between the SF and DW model, it is still difficult to classify these models. However, $\bar{p}$ is maximized for the correct model for the input, and the classification is slightly improved compared to the case of $m_{\rm DM}=2$ keV. This result is consistent with Figure~\ref{fig:k_half}. The difference between the WDM models becomes larger for more massive dark matter due to the larger difference of $k_{1/2}$, while the difference appears at smaller scales. If the WDM is a heavy particle, the classification can become more feasible with higher-resolution data. 

%
\section{\label{sec:summary}Conclusion/Discussion}
%
This paper first investigates the binary classification between the CDM and FD models with various particle masses. We conduct a suite of the hydrodynamic simulations of the CDM model and FD models and generate the images of the $\delta T_b$, which is the signal from HI. We classify the images between the CDM and FD with CNN and the power spectrum and show that CNN outperforms the power spectrum for the classifications as shown in Figure~\ref{fig:AUC_dTb}. The AUC of CNN is $\sim 0.9$ for the CDM and 2 keV FD, while the one of the power spectrum is $\sim 0.6$. In addition, we do not find the redshift dependence of the classification for both methods. At higher redshift, the difference of the power spectrum for the CDM and FD is larger, as shown in Figure~\ref{fig:pk_dTb_z}, but the number of halos is smaller, and the classifications are not improved.

We also consider the joint of the images for different redshifts. We classify the images of the CDM and FD with CNN and the power spectrum, where the machine learning architectures are shown in Figure~\ref{fig:joint-z_arch}. The results are shown in Figure~\ref{fig:AUC_dTb}, and we find the classification results for the joint redshift are improved compared to those for the single redshift in both CNN and the power spectrum cases. However, the classification between the CDM and 3 keV FD WDM is not improved even when considering the joint redshift. The power spectrum analysis reveals that information derived solely from the power spectrum is insufficient to distinguish 3 keV FD WDM from CDM. This finding underscores the superiority of CNN analysis in distinguishing different dark matter models.

Next, we consider the case of the noised images. We assume the SKA-Low observation and add the white noise to the images. Then, we classify the noised images between the CDM and FD models. The results are shown in Figure~\ref{fig:AUC_dTb_noise}. For the images with noise, CNN shows better performance than the power spectrum. For the 1 keV FD case, the AUC for CNN is 0.84 at $z=3$ while the one for the power spectrum is  0.64. Additionally, we find the redshift dependence of the results for both CNN and the power spectrum. This is because the noise at higher redshift hides more signals, as we can see in Figure~\ref{fig:noised_images}.

The classification with CNN for the 1\,keV FD is improved for the joint redshift. The AUC is 0.98 for the joint redshift, while the one is 0.84 for only $z=3$. However, the classification is not significantly improved when CNN or the power spectrum cannot distinguish the images for the single redshift image, such as in the case of the 2\,keV FD WDM.

Finally, we consider additional three WDM models: $\nu$MSM, SF, and DW production scenarios. We conduct the additional simulations for these WDM models and make images. And then, we classify the dark matter models with CNN for the joint redshift.

We consider the case that the particle mass of the WDM is 2\,keV. For the 5-class classification, CNN can distinguish the CDM, FD, and $\nu$MSM from the other dark matter models. However, CNN cannot classify the SF and DW models due to the small difference in their initial power spectra. This misclassification is not changed for the case of 4-class classification without CDM, where we consider only WDM models. Even for the binary classification between the SF and DW, CNN cannot distinguish these two models for 2keV mass. 

When we consider the WDM particle mass of 3\,keV, the classifications are degraded as shown in Figure~\ref{fig:wdm_classification_3keV} due to the smaller free-streaming length of the WDM. However, the difference between the WDM models is larger at a smaller scale when we consider heavier particles, as shown in Figure~\ref{fig:k_half}. The classification of different WDM models can be more feasible for a higher WDM mass if we have images with a high enough resolution. Our study shows that the classification between the SF and DW with the mass 3 keV, for instance, is indeed improved compared to the 2\,keV case.

Further comments on the treatment for the different dark matter production mechanisms are in order. Our approach simplified the representation of differences in production mechanisms by rescaling the warm dark matter mass within a Fermi-Dirac distribution. This methodology provides valuable insights and emphasizes the potential of 21 cm signal observations in probing warm dark matter models. However, a more rigorous examination of these scenarios is warranted for a more comprehensive exploration of dark matter properties.
Having demonstrated that different production scenarios can leave observable imprints on the 21 cm signals and, more broadly, on cosmic structure formation, the next step would be to incorporate these varied phase space distributions into the collisional terms of the Boltzmann equations. Furthermore, integrating the Boltzmann equation in a six-dimensional phase space, although computationally intensive, would be a vital future step toward a deeper understanding of dark matter characteristics \cite{2013ApJ...762..116Y,2021arXiv211015867Y,2020ApJ...904..159Y}.
Our use of machine learning techniques, which leverage image recognition and thus can capture higher-order correlation information beyond the simple two-point correlation functions, would also benefit from such a thorough analysis and is expected to provide deeper insights into the nature of dark matter.
The question of how dark matter was produced in the early Universe continues to be one of the most intriguing challenges at the crossroads of particle physics and cosmology, and further exploration of dark matter production mechanisms through the cosmological observations, including the treatment of the Boltzmann equation in six-dimensional phase space, is left for future work.

\revi{As the volume of observational data continues to grow, the need for efficient data processing becomes increasingly critical for forthcoming observations. While CNNs outperform other methodologies, their extensive consumption of computational resources limits the complexity that can be incorporated into machine learning models. This significant drawback hampers efforts to constrain dark matter models. To enhance data processing efficiency, we intend to explore alternative statistical methods or more efficient architectures, such as Graph Neural Networks (GNNs) \citep{4700287}. This exploration will be the focus of our future work.}

\begin{acknowledgments}
K.M. would like to thank the “Nagoya University Interdisciplinary Frontier Fellowship” supported by Nagoya University and JST, the establishment of university fellowships toward the creation of science technology innovation, Grant No. JPMJFS2120.
We are grateful to Volker Springel for providing the original version of \texttt{GADGET-3}, on which the \texttt{GADGET3-Osaka} code is based.
Part of the computation is performed on Cray XC50 and GPU cluster at the CfCA in NAOJ and the GPU workstation at Nagoya University.
We also acknowledge the computing support for the SQUID at the Cybermedia Center, Osaka University as part of the HPCI system Research Project (hp220044, hp230089). This work was supported in part by Center of Quantum Cosmo Theoretical Physics (NSFC grant number 12347103) (K.K.), the MEXT/JSPS KAKENHI grant No. JP19H05810, JP20H00180 (K.N.), JP21H05454, and JP21K03625 (A.N.) and the JSPS program, International Leading Research, JP22K21349 (K.N./A.N.).
\end{acknowledgments}

%

\bibliography{main}

\end{document}